\documentclass[%
 reprint,
 superscriptaddress,
 amsmath,amssymb,
 prl,
]{revtex4-1}
\usepackage{filecontents}

\usepackage{graphicx}% Include figure files
\usepackage{dcolumn}% Align table columns on decimal point
\usepackage{bm}% bold math
\usepackage{physics}
\usepackage[dvipsnames]{xcolor}
\usepackage[
    colorlinks,
    citecolor=RoyalBlue,
    linkcolor=RoyalBlue,
    urlcolor=RoyalBlue,
]{hyperref}
\usepackage{multirow}
\usepackage{mathtools}
\usepackage{verbatim}  % for comments
\usepackage[normalem]{ulem}

\begin{document}

\title{Even-order optical harmonics generated from centrosymmetric-material metasurfaces}

\author{Pavel Tonkaev}
\affiliation{Nonlinear Physics Centre, Australian National University, Canberra, ACT 2601, Australia}

\author{Fangxing Lai}
\affiliation{Ministry of Industry and Information Technology Key Lab of Micro-Nano Optoelectronic Information System, Guangdong Provincial Key Laboratory of Semiconductor Optoelectronic Materials and Intelligent Photonic Systems, Harbin Institute of Technology Shenzhen 518055, P. R. China}

\author{Sergey Kruk}
\affiliation{Nonlinear Physics Centre, Australian National University, Canberra, ACT 2601, Australia}

\author{Qinghai Song}
\affiliation{Ministry of Industry and Information Technology Key Lab of Micro-Nano Optoelectronic Information System, Guangdong Provincial Key Laboratory of Semiconductor Optoelectronic Materials and Intelligent Photonic Systems, Harbin Institute of Technology Shenzhen 518055, P. R. China}

\author{Michael Scalora}
\affiliation{Aviation and Missile Center, U.S. Army CCDC, Redstone Arsenal, Alabama 35898-5000, USA}

\author{Kirill Koshelev}
\affiliation{Nonlinear Physics Centre, Australian National University, Canberra, ACT 2601, Australia}

\author{Yuri Kivshar}
\affiliation{Nonlinear Physics Centre, Australian National University, Canberra, ACT 2601, Australia}
\email{yuri.kivshar@anu.edu.au}

\begin{abstract}
Generation of even-order optical harmonics requires noncentrosymmetric structures being conventionally observed in crystals lacking the center of inversion. In centrosymmetric systems, even-order harmonics may arise, e.g., at surfaces but such effects are usually very weak. Here we observe optical harmonics up to 4-th order generated under the normal incidence from {\it centrosymmetric} dielectric metasurfaces empowered by resonances. We design silicon metasurfaces supporting optical quasi-bound states in the continuum and guided-mode resonances, and demonstrate the enhancement of second-harmonic signals by over three orders of magnitude compared to nonresonant thin films. Under the optimal conditions, the brightness of the second harmonic approaches that of the third harmonic, and the 4th-order harmonic becomes detectable.
\end{abstract}

\maketitle

Generation of high-order optical harmonics is traditionally associated with plasmas and gases~\cite{burnett1977harmonic, ferray1988multiple}, but it has recently entered the realm of nanostructured solids~\cite{vampa2017plasmon,liu2018enhanced,shcherbakov2021generation,zhang2021enhanced,schmid2021tunable,an2021efficient,zubyuk2021resonant,zograf2022high,abbing2022extreme,zalogina2023high,jangid2024spectral}. Solid-state structures, unlike gases, enable efficient generation of even-order harmonics -- a process reliant on the absence of inversion-point symmetry. The dominant approach towards even-order harmonic generation in nanophotonics is to employ materials with noncentrosymmetric crystalline lattices~\cite{liu2018all,shcherbakov2021generation}.

In non-conductive, bulk centrosymmetric media, even-order nonlinear optical processes can be triggered by several factors, including surface nonlinearities~\cite{timbrell2018comparative}, strong external fields leading to inversion symmetry breaking~\cite{bouhelier2003near, timurdogan2017electric, lee2019electrically}, as well as by applying an external strain~\cite{schriever2010strain} or engineering internal strains by an additional overlayer~\cite{cazzanelli2012second,zhao2022second}. In the presence of free electrons in the conduction band, even-order nonlinearities may be enabled by the magnetic component of the Lorentz force~\cite{guo2019enhanced,yang2021enhanced}. Observations of second-harmonic generation (SHG) were made in various nanoscale systems made of silicon (centrosymmetric material), including membranes~\cite{scalora2019resonant} often enhanced by Fabry-Perot resonances~\cite{hallman2023harmonic}, spherical nanoparticles~\cite{makarov2017efficient} and nanowires~\cite{wiecha2015enhanced,wiecha2016origin} enhanced by Mie resonances, photonic crystal cavities~\cite{galli2010low} as well as metasurfaces~\cite{bar2019nonlinear,xu2021second,wang2024resonantly}.

However, even-order effects from centrosymmetric materials are usually very weak, in sharp contrast to odd-order nonlinearities that are driven by volume nonlinear sources~\cite{shen1986,sipe1987,shen1999} (such as third-harmonic generation). Despite early observations of fourth optical harmonics from bulk silicon in 1998~\cite{lee1998reflected}, in nanophotonics all observations of even-order nonlinear effects discussed above have been limited to the lowest, second-order harmonic.

%Unlike metals, dielectrics and undoped semiconductors have negligible free-charge densities. As a result, SHG was attributed completely to the Coulomb and the magnetic component of the Lorentz force acting on bound electrons. 

 Here, we experimentally demonstrate generation of optical harmonics up to 4th order from resonant silicon metasurfaces driven by bound states in the continuum (BIC) and guided-mode resonances (GMR). 
 We achieve the enhancement of the brightness of SHG, which under optimal resonant conditions approaches the brightness of third harmonic generation (THG). At resonance, we observe fourth harmonic generation (FHG) that is otherwise undetectable off-resonance.

\begin{figure}[h]
    \centering
    \includegraphics[width=0.9\linewidth]{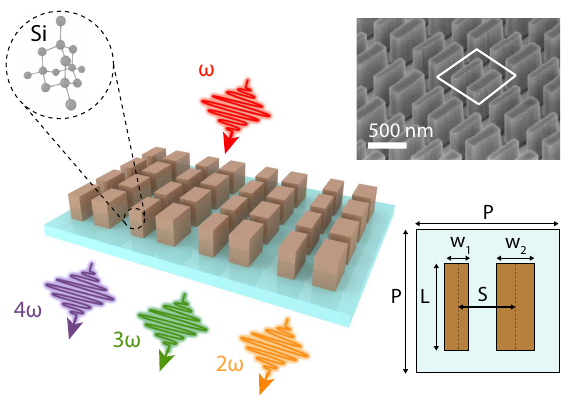}
    \caption{ \textbf{Resonant metasurface for enhanced harmonic generation}. A Si metasurface supporting BIC and GMR resonances for generating second to fourth optical harmonics. The inset shows a SEM image with a highlighted unit cell.}
    \label{fig1}
\end{figure}

 We design and fabricate metasurfaces made of amorphous silicon (a-Si) and crystalline silicon (c-Si) and compare their nonlinear properties. We characterize the metasurface's nonlinear response by pumping it in the near-IR frequency range. We observe the enhancement of over three orders of magnitude for the SHG and THG from c-Si metasurfaces driven by the GMR and BIC resonances. In contrast, the a-Si metasurface enhances SHG by two orders of magnitude. At the same time, we observe FHG from metasurfaces that support BICs. To our knowledge FHG has not been reported from silicon films or nanostructures. We also report the unexpected cubic power dependence for SHG when the pump is resonant,  whereas the slope of the SHG signal approaches square dependence when the pump is tuned off resonance.

\begin{figure*}[th!]
    \centering
    \includegraphics[width=0.8\linewidth]{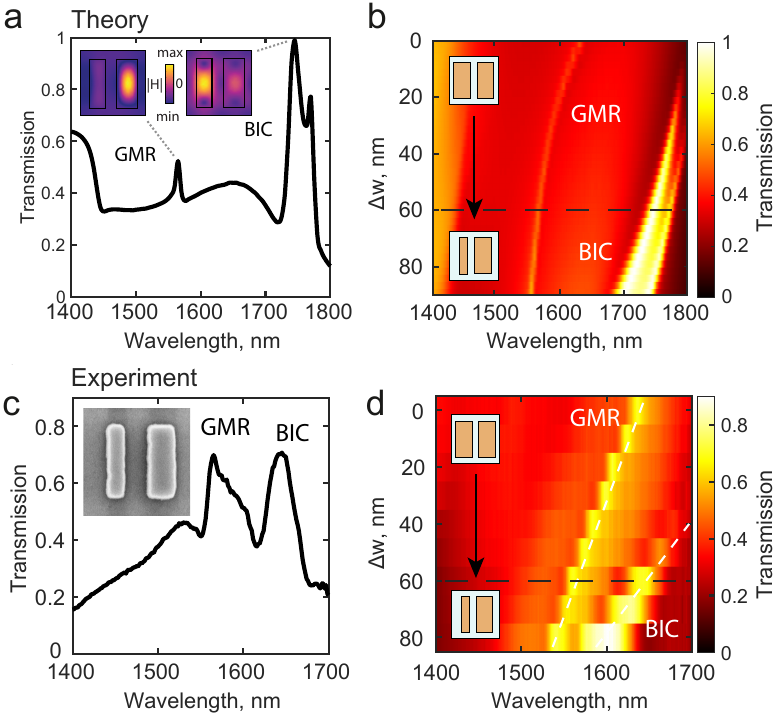}
    \caption{ \textbf{Optical properties of c-Si metasurfaces}. The numerically calculated (a) and measured (c) transmission spectrum of the metasurface with $\Delta w=60$ nm. The insets show the near-field distribution of the GMR and BIC modes, and an SEM image of the unit cell. Calculated (b) and measured (d) transmission spectra depending on the change of the asymmetric parameter $\Delta$ from an asymmetric case of 80 nm to a symmetric case of 0 nm. The black dashed lines correspond to the metasurface with $\Delta w=60$ nm.}
    \label{fig2}
\end{figure*}

The silicon metasurfaces are designed and fabricated to support both GMR and BIC resonances at the pump wavelength in the near-IR spectral range, with a-Si on glass and c-Si on sapphire. A schematic of the metasurfaces is shown in Fig.~\ref{fig1}. They are composed of a square lattice of asymmetric meta-atoms consisting of bar dimers with different widths. The metasurfaces are pumped in the region from 1.4 $\mu$m to 1.75 $\mu$m, with harmonics generated in the near-infrared and visible ranges. The inset provides the scanning electron microscope (SEM) image of c-Si metasurface with the unit cell highlighted. The geometrical parameters of the unit cell are bar height $h$=600 nm, $P$=830 nm, $L$=630 nm, $S$=380 nm, $w_2$ = 260 nm and $w_1$ is varied from 180 nm to 260 nm. Parameter $\Delta w =w_2-w_1$  is considered as an asymmetry parameter.

First, we characterize the linear transmission of c-Si metasurfaces in numerical calculations for incident polarization along the long side of the bars.  We use the finite-element-method solver in COMSOL Multiphysics in the frequency domain. All calculations were realized for a metasurface placed on a semi-infinite substrate surrounded by a perfectly matched layer mimicking an infinite region. The simulation area is the unit cell extended to an infinite metasurface by using the Bloch boundary conditions. The material properties are taken from Ref.~\cite{li1980refractive}. The transmission spectrum for $\Delta w=60$~nm in the wavelength range from 1400 to 1800 nm and transmission map for $\Delta w$ from $0$~nm to $90$~nm are shown in Figs.~\ref{fig2}a,b, respectively. The resonant features can be associated with three distinctive resonances, that have wavelengths of $1580$~nm, $1730$~nm and $1780$~nm for $\Delta w=60$~nm, respectively. We focus on two modes at $1580$~nm and $1730$~nm, and identify their nature via calculating near-field distributions in the eigenmode solver in COMSOL Multiphysics, shown in the inset of the Fig.~\ref{fig2}a for $\Delta w=60$~nm. We can see that the mode at $1730$~nm is a magnetic dipolar BIC resonance because of its distinctive asymmetric field distribution, and the mode at $1580$~nm is a magnetic dipolar GMR mode. In both modes, the volume-averaged enhancement of the local field intensity of approximately 7 times is observed (see Figure S1 in~\cite{noteSI}).

We then measure the transmission of the c-Si metasurfaces with polarization along the long side of the bars unit cell. The transmission spectrum has two Fano-shaped features corresponding to GMR and BIC resonances. Figure~\ref{fig2}c shows the transmission spectrum for the metasurfaces with $\Delta w=60$~nm. The inset shows the SEM image of the metasurface unit cell. Decreasing the asymmetry parameter causes both supporting resonances to display a red shift. Figure~\ref{fig2}d shows the transmission spectra for c-Si metasurfaces with asymmetry parameters ranging from 0 nm to 80 nm. The BIC resonance becomes less visible with decreasing $\Delta w$, whereas the GMR mode does not change in intensity. The behaviour obtained has a good agreement with theoretical prediction, which is shown in Fig.~\ref{fig2}b.  In the case of a-Si, the spectral position of the modes is shifted to shorter wavelengths, but both modes are preserved (see Figure S2 in~\cite{noteSI}). 

The nature of BIC and GMR modes is numerically and experimentally confirmed with their behaviour for $\Delta w = 0$. More specific, the BIC mode disappears from the spectrum, while the linewidth of the GMR mode barely changes with the change of $\Delta w$, as can be seen in Figs.~\ref{fig2}b,d. We note that the measured resonant positions are red shifted with respect to the simulation results, and the third mode observed in numerical transmission data is not visible in the experiments. These differences can be attributed to the tolerances of fabrication and design. Also, the mode linewidth appears to be larger in the experiment. This is due to the excitation source in experiment deviating from a plane wave due to the focusing objective and a range of modes with oblique in-plane k-vectors is excited in the vicinity of the normal incidence.

\begin{figure*}[th!]
    \centering
    \includegraphics[width=0.9\linewidth]{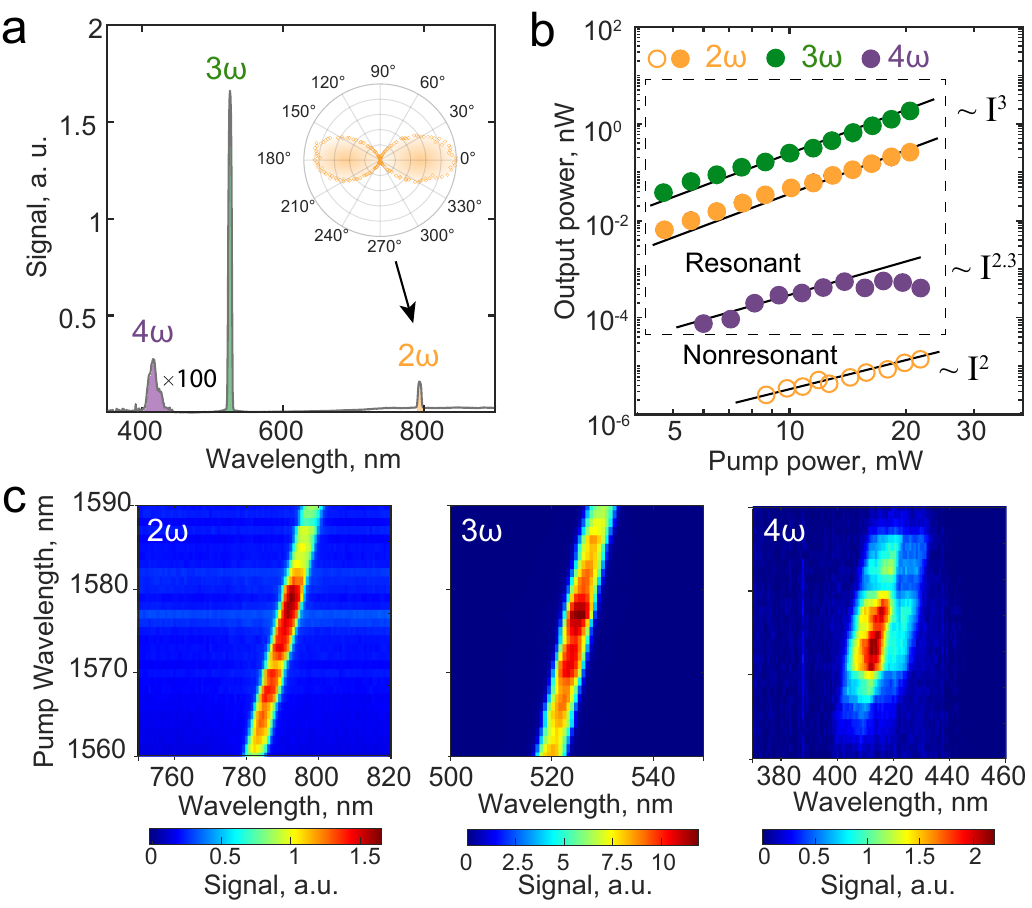}
    \caption{ \textbf{Experimental harmonic generation}. (a) Spectra of harmonic generation from a-Si metasurface with $\Delta w=60$ nm supporting BIC resonance. Fourth harmonic signal is multiplied by 100. Angular dependence of second harmonic generation on pump polarization is shown on the inset with orange dots. (b) Power-power dependence for second, third and fourth harmonics from the a-Si metasurface supporting BIC resonance, and second harmonic from the metasurface pumped nonresonantly. In the case of a resonant pump, both  SHG and THG have a slope of 3, whereas outside the resonance the second harmonic has a slope of 2. The slope for the FHG curve equals to 2.3 for low pump intensity and decreases with the pump power increase. (c) Harmonic spectra generated by the a-Si metasurface with $\Delta w=60$ nm for various pump wavelengths around BIC resonance. The integration time for FHG is a hundred times longer compared to SHG and THG. The fourth harmonic is observed only in the vicinity of the resonance.}
    \label{fig3}
\end{figure*}

To measure harmonic generation, we used a near-IR laser with a pulse duration of 5 ps, a repetition rate of 5.14 MHz, and a tunable wavelength in the range from 1.4 $\mu$m to 1.75 $\mu$m focused into a spot with a diameter of 25 $\mu m$. The signals are collected in transmission by an objective lens with NA 0.42 (see Figure S3 in~\cite{noteSI} for more details). Figure~\ref{fig3}a shows the spectrum of harmonic generation from the a-Si metasurface with $\Delta$w = 60 nm pumped resonantly at the wavelength of 1575 nm and with the peak power density of 3 GW$\cdot \textrm{cm}^{-2}$. A SHG signal is observable at 788 nm, while THG is observed at 525 nm. We remark that the magnitude of the SHG signal is one order of magnitude smaller than the magnitude of the THG signal. While this may seem unusual, the relative magnitudes of the generated harmonics are sensitive to pump tuning with respect to resonance peak, as well as material dispersion and polarization. Under ordinary circumstances, in a cavity environment it is also possible for SHG maxima to be shifted relative to THG maxima. Figure~\ref{fig3}a inset demonstrates the normalized SHG signal depends on the angle of pump polarization, where 0$^\circ$ corresponds to polarisation coinciding with the longest side of the bars. In the case of pump polarization orthogonal to the bars, the SH signal is significantly smaller. The same pattern is observed for THG (see Figure S4 in~\cite{noteSI}). The polarimetry of the THG and SHG signals reveals that both harmonics pumped at the BIC resonance preserve the polarisation of the excitation and the degrees of polarization are 0.8 and 0.7, respectively. We note that FHG can also be observed in the spectrum at the wavelength of 415 nm. Figure~\ref{fig3}c shows the harmonic spectra recorded for the various pump wavelengths around BIC resonance.  The FHG is observed only in the vicinity of the BIC resonance, whereas the SHG and THG are observed outside the resonant pump. 

\begin{figure*}[th]
    \centering
    \includegraphics[width=0.99\linewidth]{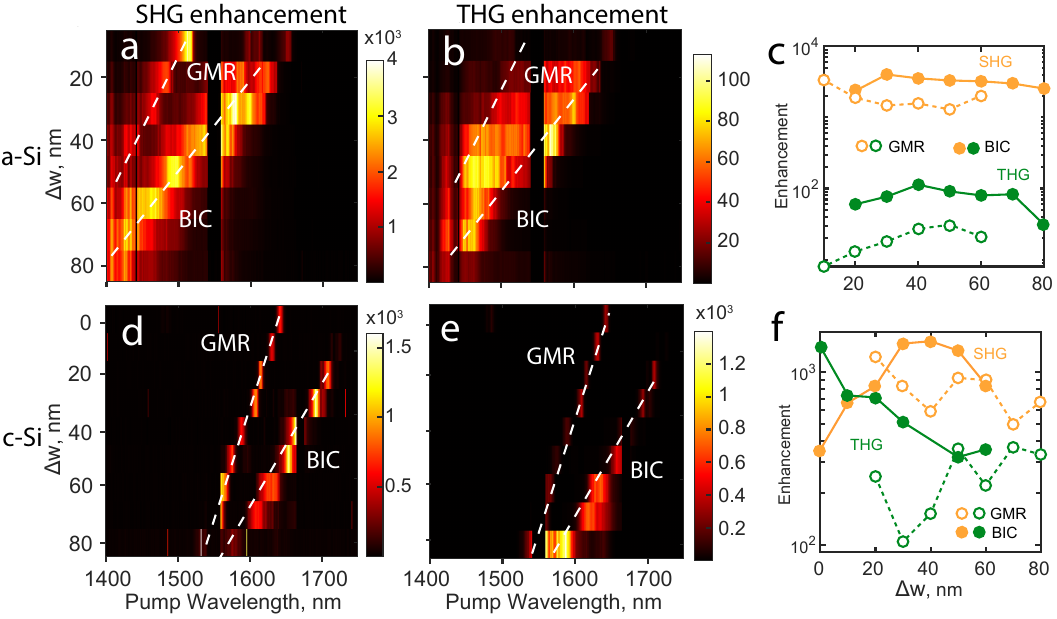}
    \caption{ \textbf{Harmonic enhancement for a-Si and c-Si metasurfaces}. Second harmonic enhancement from (a) a-Si and (d) c-Si dependence on asymmetric parameter and pump wavelength. (c) Maximum of second and third harmonics enhancement for different modes depending on the asymmetric parameter for a-Si metasurface.  Third harmonic generation enhancement from (b) a-Si and (e) c-Si dependence on asymmetric parameter and pump wavelength. (f) Maximum of second and third harmonics enhancement for different modes depending on the asymmetric parameter for c-Si metasurface.}
    \label{fig4}
\end{figure*}

We also record the power dependence of SHG, THG and FHG pumped resonantly at the wavelength of 1636 nm, which corresponds to the BIC resonance of the metasurface. The dependencies for SHG, THG and FHG are indicated in Figure~\ref{fig3}b with orange, green and purple circles, respectively. Both THG and SHG curves agree well with the cubic power law. The results recorded for the same metasurface, but pumped at a wavelength of 1470 nm, which excites the GMR mode of the structure, exhibit similar behaviour. The slope of the FHG curve is less than the expected quadratic dependence, and it equals 2.3 for low pump intensity and decreases as pump power increases. When the pump is nonresonant, the situation changes for SHG. The signal shown in Figure~\ref{fig3}b is denoted by the hollow circles, and shows good agreement with the quadratic power law. In contrast, THG preserves the same slope of 3 even when the pump is tuned outside the resonance (see Figure S5 in~\cite{noteSI}). The results for the reference c-Si and a-Si unstructured films also show slopes 2 and 3 for the SHG and THG, respectively. 
The generation of the second harmonic with a slope larger than two has previously been observed in semiconductors~\cite{fomenko2001nonquadratic, wang1998coupled} and hybrid structures~\cite{wen2018plasmonic, sun2023all}. That behaviour is caused by the generation of the static electric field by the high-intensity laser beam.

To study the dependence of the SHG and THG on asymmetry parameter and pump wavelength, we measure the harmonic signal from the metasurfaces having different asymmetries, and pump wavelengths from 1400 nm to 1750 nm. The maximum SHG and THG values are determined and normalized by the harmonic signal generated by a similar silicon film. The enhancement values obtained for a-Si metasurface are shown in Figure~\ref{fig4}(a,b). The maximum enhancement for SHG is three orders of magnitude, whereas it is about two orders of magnitudes for THG. The BIC mode is clearly observed in both second and third harmonics enhancements maps, and is in good agreement with mode positions in the linear spectra. Figure~\ref{fig4}c sums up the enhancement for SHG (orange circles) and THG (green circles). Both SHG and THG enhanced by the BIC tend to be larger than the signals enhanced by the GMR modes. 

For c-Si metasurfaces the positions of the enhancement are shifted compared to the a-Si metasurfaces, similar to the linear spectra features shift. Figures~\ref{fig4}(d,e) demonstrate the enhancement from c-Si metasurfaces depending on asymmetry parameter and pump wavelength. In contrast with a-Si metasurfaces, both modes are clearly distinguishable for SHG and THG enhancement maps. SH and TH signal enhancements are about three orders of magnitude. SH enhancement is compared to the best case scenario where the wave is incident at approximately 70 degrees ~\cite{hallman2023harmonic}. A decrease in the asymmetry parameter causes the decrease of linewidth for the BIC mode and its eventual disappearance from the spectrum for the fully symmetric case. The enhancement of SHG and THG are summarized in Figure~\ref{fig4}f by the orange and green circles, respectively. We note that meanwhile, the SHG enhancement is in the same order of magnitude for a-Si and c-Si metasurfaces, and the SHG signal from the c-Si metasurfaces is one order of magnitude lower than the SHG from the a-Si metasurface (see Figure S6 in~\cite{noteSI})

As noted earlier, we also observe the fourth harmonic generation. Since a-Si metasurfaces have modes at around 1350 nm for $\Delta w=80$ nm to 1650 nm for $\Delta w=10$ nm according to the linear transmission spectra, the spectral position of the fourth harmonic is located either just outside or at the edge of the spectrometer's range. Thus, it can be detected only for the BIC resonance for the a-Si metasurfaces with size parameters that yield near-symmetric structures. However, for those size parameters, the BIC resonance is either weak or not observed. In contrast, c-Si metasurfaces have modes that are spectrally redshifted and enable the observation of the fourth harmonic within the spectrometer spectral range. However, the harmonics themselves are weaker for the c-Si, which complicates their observation.

There are several mechanisms proposed to explain the symmetry breaking in silicon that leads to even harmonic generation. Since there is no proper band structure at the interface, nonlocal effects contribute to surface and quadrupolar bulk SHG~\cite{tancogne2016ab}. An intense pump laser pulse causes weak direct-current static electric field that affects the phases of the optical field half-cycles and breaks the symmetry, producing even harmonics in silicon~\cite{apostolova2021high,ovchinnikov2019second}. Inhomogeneous laser fields in the vicinity of plasmonic resonances can also produce even harmonics~\cite{ciappina2012high}. This scenario might be relevant for strongly excited semiconductors, but this issue is beyond the scope of this work.

Even harmonics may be triggered by a combination of surface and magnetic terms~\cite{hallman2023harmonic,scalora2019resonant, mukhopadhyay2023three}. For example, the bound electron term $\left(\mathbf{P}_\omega \cdot \nabla\right) \mathbf{E}_\omega$  and the magnetic term $\mathbf{J}_\omega \times \mathbf{H}_\omega$  are the dominant second harmonic sources. Here, polarization, current, and fields are to be interpreted as envelope functions at the given frequency. Similarly, fourth harmonic nonlinear surface and magnetic sources may also be derived and written as $\left(\mathbf{P}_{2 \omega} \cdot \nabla\right) \mathbf{E}_{2 \omega}$  and $\mathbf{J}_{2 \omega} \times \mathbf{H}_{2 \omega}$. However, efficient third harmonic generation activates additional surface and magnetic terms that are proportional to  $\left(\mathbf{P}_{3 \omega} \cdot \nabla\right) \mathbf{E}_\omega$, $\mathbf{J}_{3 \omega} \times \mathbf{H}_\omega$, and $\mathbf{J}_\omega \times \mathbf{H}_{3 \omega}$, as well as fourth order bulk terms proportional to $\chi^{(3)} E_{3 \omega} E_\omega$.  Therefore, there is some expectation that fourth harmonic generation will fall within the range of parameters that determine a combination of second and third harmonic generation. Currently, while the theory can provide good guidance, we do not have a clear theoretical framework for the quantitative aspects of our observations.  

In summary, we have studied optical harmonic generation from metasurfaces made of amorphous and crystalline silicon. We have observed the generation of the second and third harmonics enhanced by at least three orders of magnitude via GMR and BIC resonances in the near-IR range. This enabled significant generation of the fourth harmonics at the BIC resonance. We have observed the deviation in the power laws for both even harmonics at the BIC resonance. We believe that our results reveal that dielectric metasurfaces made of centrosymmetric materials can exhibit generally unexpectedly strong even-order nonlinear effects when they are empowered by high-$Q$ optical resonances associated with the bound states in the continuum.

\begin{acknowledgements}
The authors thank C.~Cojocaru, S.A.~Maier, N.C. Panoiu, J. Trull, and M.A. Vincenti for useful discussions and suggestions.  This work was supported by the Australian Research Council (Grants DP210101292 and DE210100679), the International Technology Center Indo-Pacific (ITC IPAC) via Army Research Office (contract FA520923C0023), and National Natural Science Foundation of China (Grants 12334016, 12025402, and 12261131500).
\end{acknowledgements}


\begin{thebibliography}{47}%
\makeatletter
\providecommand \@ifxundefined [1]{%
 \@ifx{#1\undefined}
}%
\providecommand \@ifnum [1]{%
 \ifnum #1\expandafter \@firstoftwo
 \else \expandafter \@secondoftwo
 \fi
}%
\providecommand \@ifx [1]{%
 \ifx #1\expandafter \@firstoftwo
 \else \expandafter \@secondoftwo
 \fi
}%
\providecommand \natexlab [1]{#1}%
\providecommand \enquote  [1]{``#1''}%
\providecommand \bibnamefont  [1]{#1}%
\providecommand \bibfnamefont [1]{#1}%
\providecommand \citenamefont [1]{#1}%
\providecommand \href@noop [0]{\@secondoftwo}%
\providecommand \href [0]{\begingroup \@sanitize@url \@href}%
\providecommand \@href[1]{\@@startlink{#1}\@@href}%
\providecommand \@@href[1]{\endgroup#1\@@endlink}%
\providecommand \@sanitize@url [0]{\catcode `\\12\catcode `\$12\catcode `\&12\catcode `\#12\catcode `\^12\catcode `\_12\catcode `\%12\relax}%
\providecommand \@@startlink[1]{}%
\providecommand \@@endlink[0]{}%
\providecommand \url  [0]{\begingroup\@sanitize@url \@url }%
\providecommand \@url [1]{\endgroup\@href {#1}{\urlprefix }}%
\providecommand \urlprefix  [0]{URL }%
\providecommand \Eprint [0]{\href }%
\providecommand \doibase [0]{http://dx.doi.org/}%
\providecommand \selectlanguage [0]{\@gobble}%
\providecommand \bibinfo  [0]{\@secondoftwo}%
\providecommand \bibfield  [0]{\@secondoftwo}%
\providecommand \translation [1]{[#1]}%
\providecommand \BibitemOpen [0]{}%
\providecommand \bibitemStop [0]{}%
\providecommand \bibitemNoStop [0]{.\EOS\space}%
\providecommand \EOS [0]{\spacefactor3000\relax}%
\providecommand \BibitemShut  [1]{\csname bibitem#1\endcsname}%
\let\auto@bib@innerbib\@empty
%</preamble>
\bibitem [{\citenamefont {Burnett}\ \emph {et~al.}(1977)\citenamefont {Burnett}, \citenamefont {Baldis}, \citenamefont {Richardson},\ and\ \citenamefont {Enright}}]{burnett1977harmonic}%
  \BibitemOpen
  \bibfield  {author} {\bibinfo {author} {\bibfnamefont {N.}~\bibnamefont {Burnett}}, \bibinfo {author} {\bibfnamefont {H.}~\bibnamefont {Baldis}}, \bibinfo {author} {\bibfnamefont {M.}~\bibnamefont {Richardson}}, \ and\ \bibinfo {author} {\bibfnamefont {G.}~\bibnamefont {Enright}},\ }\href@noop {} {\bibfield  {journal} {\bibinfo  {journal} {Applied Physics Letters}\ }\textbf {\bibinfo {volume} {31}},\ \bibinfo {pages} {172} (\bibinfo {year} {1977})}\BibitemShut {NoStop}%
\bibitem [{\citenamefont {Ferray}\ \emph {et~al.}(1988)\citenamefont {Ferray}, \citenamefont {L'Huillier}, \citenamefont {Li}, \citenamefont {Lompre}, \citenamefont {Mainfray},\ and\ \citenamefont {Manus}}]{ferray1988multiple}%
  \BibitemOpen
  \bibfield  {author} {\bibinfo {author} {\bibfnamefont {M.}~\bibnamefont {Ferray}}, \bibinfo {author} {\bibfnamefont {A.}~\bibnamefont {L'Huillier}}, \bibinfo {author} {\bibfnamefont {X.}~\bibnamefont {Li}}, \bibinfo {author} {\bibfnamefont {L.}~\bibnamefont {Lompre}}, \bibinfo {author} {\bibfnamefont {G.}~\bibnamefont {Mainfray}}, \ and\ \bibinfo {author} {\bibfnamefont {C.}~\bibnamefont {Manus}},\ }\href@noop {} {\bibfield  {journal} {\bibinfo  {journal} {Journal of Physics B: Atomic, Molecular and Optical Physics}\ }\textbf {\bibinfo {volume} {21}},\ \bibinfo {pages} {L31} (\bibinfo {year} {1988})}\BibitemShut {NoStop}%
\bibitem [{\citenamefont {Vampa}\ \emph {et~al.}(2017)\citenamefont {Vampa}, \citenamefont {Ghamsari}, \citenamefont {Siadat~Mousavi}, \citenamefont {Hammond}, \citenamefont {Olivieri}, \citenamefont {Lisicka-Skrek}, \citenamefont {Naumov}, \citenamefont {Villeneuve}, \citenamefont {Staudte}, \citenamefont {Berini} \emph {et~al.}}]{vampa2017plasmon}%
  \BibitemOpen
  \bibfield  {author} {\bibinfo {author} {\bibfnamefont {G.}~\bibnamefont {Vampa}}, \bibinfo {author} {\bibfnamefont {B.}~\bibnamefont {Ghamsari}}, \bibinfo {author} {\bibfnamefont {S.}~\bibnamefont {Siadat~Mousavi}}, \bibinfo {author} {\bibfnamefont {T.}~\bibnamefont {Hammond}}, \bibinfo {author} {\bibfnamefont {A.}~\bibnamefont {Olivieri}}, \bibinfo {author} {\bibfnamefont {E.}~\bibnamefont {Lisicka-Skrek}}, \bibinfo {author} {\bibfnamefont {A.~Y.}\ \bibnamefont {Naumov}}, \bibinfo {author} {\bibfnamefont {D.}~\bibnamefont {Villeneuve}}, \bibinfo {author} {\bibfnamefont {A.}~\bibnamefont {Staudte}}, \bibinfo {author} {\bibfnamefont {P.}~\bibnamefont {Berini}},  \emph {et~al.},\ }\href@noop {} {\bibfield  {journal} {\bibinfo  {journal} {Nature Physics}\ }\textbf {\bibinfo {volume} {13}},\ \bibinfo {pages} {659} (\bibinfo {year} {2017})}\BibitemShut {NoStop}%
\bibitem [{\citenamefont {Liu}\ \emph {et~al.}(2018{\natexlab{a}})\citenamefont {Liu}, \citenamefont {Guo}, \citenamefont {Vampa}, \citenamefont {Zhang}, \citenamefont {Sarmiento}, \citenamefont {Xiao}, \citenamefont {Bucksbaum}, \citenamefont {Vu{\v{c}}kovi{\'c}}, \citenamefont {Fan},\ and\ \citenamefont {Reis}}]{liu2018enhanced}%
  \BibitemOpen
  \bibfield  {author} {\bibinfo {author} {\bibfnamefont {H.}~\bibnamefont {Liu}}, \bibinfo {author} {\bibfnamefont {C.}~\bibnamefont {Guo}}, \bibinfo {author} {\bibfnamefont {G.}~\bibnamefont {Vampa}}, \bibinfo {author} {\bibfnamefont {J.~L.}\ \bibnamefont {Zhang}}, \bibinfo {author} {\bibfnamefont {T.}~\bibnamefont {Sarmiento}}, \bibinfo {author} {\bibfnamefont {M.}~\bibnamefont {Xiao}}, \bibinfo {author} {\bibfnamefont {P.~H.}\ \bibnamefont {Bucksbaum}}, \bibinfo {author} {\bibfnamefont {J.}~\bibnamefont {Vu{\v{c}}kovi{\'c}}}, \bibinfo {author} {\bibfnamefont {S.}~\bibnamefont {Fan}}, \ and\ \bibinfo {author} {\bibfnamefont {D.~A.}\ \bibnamefont {Reis}},\ }\href@noop {} {\bibfield  {journal} {\bibinfo  {journal} {Nature Physics}\ }\textbf {\bibinfo {volume} {14}},\ \bibinfo {pages} {1006} (\bibinfo {year} {2018}{\natexlab{a}})}\BibitemShut {NoStop}%
\bibitem [{\citenamefont {Shcherbakov}\ \emph {et~al.}(2021)\citenamefont {Shcherbakov}, \citenamefont {Zhang}, \citenamefont {Tripepi}, \citenamefont {Sartorello}, \citenamefont {Talisa}, \citenamefont {AlShafey}, \citenamefont {Fan}, \citenamefont {Twardowski}, \citenamefont {Krivitsky}, \citenamefont {Kuznetsov} \emph {et~al.}}]{shcherbakov2021generation}%
  \BibitemOpen
  \bibfield  {author} {\bibinfo {author} {\bibfnamefont {M.~R.}\ \bibnamefont {Shcherbakov}}, \bibinfo {author} {\bibfnamefont {H.}~\bibnamefont {Zhang}}, \bibinfo {author} {\bibfnamefont {M.}~\bibnamefont {Tripepi}}, \bibinfo {author} {\bibfnamefont {G.}~\bibnamefont {Sartorello}}, \bibinfo {author} {\bibfnamefont {N.}~\bibnamefont {Talisa}}, \bibinfo {author} {\bibfnamefont {A.}~\bibnamefont {AlShafey}}, \bibinfo {author} {\bibfnamefont {Z.}~\bibnamefont {Fan}}, \bibinfo {author} {\bibfnamefont {J.}~\bibnamefont {Twardowski}}, \bibinfo {author} {\bibfnamefont {L.~A.}\ \bibnamefont {Krivitsky}}, \bibinfo {author} {\bibfnamefont {A.~I.}\ \bibnamefont {Kuznetsov}},  \emph {et~al.},\ }\href@noop {} {\bibfield  {journal} {\bibinfo  {journal} {Nature communications}\ }\textbf {\bibinfo {volume} {12}},\ \bibinfo {pages} {4185} (\bibinfo {year} {2021})}\BibitemShut {NoStop}%
\bibitem [{\citenamefont {Zhang}\ \emph {et~al.}(2021)\citenamefont {Zhang}, \citenamefont {Tu}, \citenamefont {Wu}, \citenamefont {Lyu}, \citenamefont {Zhao},\ and\ \citenamefont {Yuan}}]{zhang2021enhanced}%
  \BibitemOpen
  \bibfield  {author} {\bibinfo {author} {\bibfnamefont {D.}~\bibnamefont {Zhang}}, \bibinfo {author} {\bibfnamefont {Y.}~\bibnamefont {Tu}}, \bibinfo {author} {\bibfnamefont {H.}~\bibnamefont {Wu}}, \bibinfo {author} {\bibfnamefont {Z.}~\bibnamefont {Lyu}}, \bibinfo {author} {\bibfnamefont {Z.}~\bibnamefont {Zhao}}, \ and\ \bibinfo {author} {\bibfnamefont {J.}~\bibnamefont {Yuan}},\ }in\ \href@noop {} {\emph {\bibinfo {booktitle} {2021 46th International Conference on Infrared, Millimeter and Terahertz Waves (IRMMW-THz)}}}\ (\bibinfo {organization} {IEEE},\ \bibinfo {year} {2021})\ pp.\ \bibinfo {pages} {1--2}\BibitemShut {NoStop}%
\bibitem [{\citenamefont {Schmid}\ \emph {et~al.}(2021)\citenamefont {Schmid}, \citenamefont {Weigl}, \citenamefont {Gr{\"o}ssing}, \citenamefont {Junk}, \citenamefont {Gorini}, \citenamefont {Schlauderer}, \citenamefont {Ito}, \citenamefont {Meierhofer}, \citenamefont {Hofmann}, \citenamefont {Afanasiev} \emph {et~al.}}]{schmid2021tunable}%
  \BibitemOpen
  \bibfield  {author} {\bibinfo {author} {\bibfnamefont {C.~P.}\ \bibnamefont {Schmid}}, \bibinfo {author} {\bibfnamefont {L.}~\bibnamefont {Weigl}}, \bibinfo {author} {\bibfnamefont {P.}~\bibnamefont {Gr{\"o}ssing}}, \bibinfo {author} {\bibfnamefont {V.}~\bibnamefont {Junk}}, \bibinfo {author} {\bibfnamefont {C.}~\bibnamefont {Gorini}}, \bibinfo {author} {\bibfnamefont {S.}~\bibnamefont {Schlauderer}}, \bibinfo {author} {\bibfnamefont {S.}~\bibnamefont {Ito}}, \bibinfo {author} {\bibfnamefont {M.}~\bibnamefont {Meierhofer}}, \bibinfo {author} {\bibfnamefont {N.}~\bibnamefont {Hofmann}}, \bibinfo {author} {\bibfnamefont {D.}~\bibnamefont {Afanasiev}},  \emph {et~al.},\ }\href@noop {} {\bibfield  {journal} {\bibinfo  {journal} {Nature}\ }\textbf {\bibinfo {volume} {593}},\ \bibinfo {pages} {385} (\bibinfo {year} {2021})}\BibitemShut {NoStop}%
\bibitem [{\citenamefont {An}\ and\ \citenamefont {Kim}(2021)}]{an2021efficient}%
  \BibitemOpen
  \bibfield  {author} {\bibinfo {author} {\bibfnamefont {J.-K.}\ \bibnamefont {An}}\ and\ \bibinfo {author} {\bibfnamefont {K.-H.}\ \bibnamefont {Kim}},\ }\href@noop {} {\bibfield  {journal} {\bibinfo  {journal} {Optics \& Laser Technology}\ }\textbf {\bibinfo {volume} {135}},\ \bibinfo {pages} {106702} (\bibinfo {year} {2021})}\BibitemShut {NoStop}%
\bibitem [{\citenamefont {Zubyuk}\ \emph {et~al.}(2021)\citenamefont {Zubyuk}, \citenamefont {Carletti}, \citenamefont {Shcherbakov},\ and\ \citenamefont {Kruk}}]{zubyuk2021resonant}%
  \BibitemOpen
  \bibfield  {author} {\bibinfo {author} {\bibfnamefont {V.}~\bibnamefont {Zubyuk}}, \bibinfo {author} {\bibfnamefont {L.}~\bibnamefont {Carletti}}, \bibinfo {author} {\bibfnamefont {M.}~\bibnamefont {Shcherbakov}}, \ and\ \bibinfo {author} {\bibfnamefont {S.}~\bibnamefont {Kruk}},\ }\href@noop {} {\bibfield  {journal} {\bibinfo  {journal} {APL Materials}\ }\textbf {\bibinfo {volume} {9}} (\bibinfo {year} {2021})}\BibitemShut {NoStop}%
\bibitem [{\citenamefont {Zograf}\ \emph {et~al.}(2022)\citenamefont {Zograf}, \citenamefont {Koshelev}, \citenamefont {Zalogina}, \citenamefont {Korolev}, \citenamefont {Hollinger}, \citenamefont {Choi}, \citenamefont {Zuerch}, \citenamefont {Spielmann}, \citenamefont {Luther-Davies}, \citenamefont {Kartashov} \emph {et~al.}}]{zograf2022high}%
  \BibitemOpen
  \bibfield  {author} {\bibinfo {author} {\bibfnamefont {G.}~\bibnamefont {Zograf}}, \bibinfo {author} {\bibfnamefont {K.}~\bibnamefont {Koshelev}}, \bibinfo {author} {\bibfnamefont {A.}~\bibnamefont {Zalogina}}, \bibinfo {author} {\bibfnamefont {V.}~\bibnamefont {Korolev}}, \bibinfo {author} {\bibfnamefont {R.}~\bibnamefont {Hollinger}}, \bibinfo {author} {\bibfnamefont {D.-Y.}\ \bibnamefont {Choi}}, \bibinfo {author} {\bibfnamefont {M.}~\bibnamefont {Zuerch}}, \bibinfo {author} {\bibfnamefont {C.}~\bibnamefont {Spielmann}}, \bibinfo {author} {\bibfnamefont {B.}~\bibnamefont {Luther-Davies}}, \bibinfo {author} {\bibfnamefont {D.}~\bibnamefont {Kartashov}},  \emph {et~al.},\ }\href@noop {} {\bibfield  {journal} {\bibinfo  {journal} {ACS Photonics}\ }\textbf {\bibinfo {volume} {9}},\ \bibinfo {pages} {567} (\bibinfo {year} {2022})}\BibitemShut {NoStop}%
\bibitem [{\citenamefont {Abbing}\ \emph {et~al.}(2022)\citenamefont {Abbing}, \citenamefont {Kolkowski}, \citenamefont {Zhang}, \citenamefont {Campi}, \citenamefont {L{\"o}tgering}, \citenamefont {Koenderink},\ and\ \citenamefont {Kraus}}]{abbing2022extreme}%
  \BibitemOpen
  \bibfield  {author} {\bibinfo {author} {\bibfnamefont {S.~D.~R.}\ \bibnamefont {Abbing}}, \bibinfo {author} {\bibfnamefont {R.}~\bibnamefont {Kolkowski}}, \bibinfo {author} {\bibfnamefont {Z.-Y.}\ \bibnamefont {Zhang}}, \bibinfo {author} {\bibfnamefont {F.}~\bibnamefont {Campi}}, \bibinfo {author} {\bibfnamefont {L.}~\bibnamefont {L{\"o}tgering}}, \bibinfo {author} {\bibfnamefont {A.~F.}\ \bibnamefont {Koenderink}}, \ and\ \bibinfo {author} {\bibfnamefont {P.~M.}\ \bibnamefont {Kraus}},\ }\href@noop {} {\bibfield  {journal} {\bibinfo  {journal} {Physical review letters}\ }\textbf {\bibinfo {volume} {128}},\ \bibinfo {pages} {223902} (\bibinfo {year} {2022})}\BibitemShut {NoStop}%
\bibitem [{\citenamefont {Zalogina}\ \emph {et~al.}(2023)\citenamefont {Zalogina}, \citenamefont {Carletti}, \citenamefont {Rudenko}, \citenamefont {Moloney}, \citenamefont {Tripathi}, \citenamefont {Lee}, \citenamefont {Shadrivov}, \citenamefont {Park}, \citenamefont {Kivshar},\ and\ \citenamefont {Kruk}}]{zalogina2023high}%
  \BibitemOpen
  \bibfield  {author} {\bibinfo {author} {\bibfnamefont {A.}~\bibnamefont {Zalogina}}, \bibinfo {author} {\bibfnamefont {L.}~\bibnamefont {Carletti}}, \bibinfo {author} {\bibfnamefont {A.}~\bibnamefont {Rudenko}}, \bibinfo {author} {\bibfnamefont {J.~V.}\ \bibnamefont {Moloney}}, \bibinfo {author} {\bibfnamefont {A.}~\bibnamefont {Tripathi}}, \bibinfo {author} {\bibfnamefont {H.-C.}\ \bibnamefont {Lee}}, \bibinfo {author} {\bibfnamefont {I.}~\bibnamefont {Shadrivov}}, \bibinfo {author} {\bibfnamefont {H.-G.}\ \bibnamefont {Park}}, \bibinfo {author} {\bibfnamefont {Y.}~\bibnamefont {Kivshar}}, \ and\ \bibinfo {author} {\bibfnamefont {S.~S.}\ \bibnamefont {Kruk}},\ }\href@noop {} {\bibfield  {journal} {\bibinfo  {journal} {Science Advances}\ }\textbf {\bibinfo {volume} {9}},\ \bibinfo {pages} {eadg2655} (\bibinfo {year} {2023})}\BibitemShut {NoStop}%
\bibitem [{\citenamefont {Jangid}\ \emph {et~al.}(2024)\citenamefont {Jangid}, \citenamefont {Richter}, \citenamefont {Tseng}, \citenamefont {Sinev}, \citenamefont {Kruk}, \citenamefont {Altug},\ and\ \citenamefont {Kivshar}}]{jangid2024spectral}%
  \BibitemOpen
  \bibfield  {author} {\bibinfo {author} {\bibfnamefont {P.}~\bibnamefont {Jangid}}, \bibinfo {author} {\bibfnamefont {F.~U.}\ \bibnamefont {Richter}}, \bibinfo {author} {\bibfnamefont {M.~L.}\ \bibnamefont {Tseng}}, \bibinfo {author} {\bibfnamefont {I.}~\bibnamefont {Sinev}}, \bibinfo {author} {\bibfnamefont {S.}~\bibnamefont {Kruk}}, \bibinfo {author} {\bibfnamefont {H.}~\bibnamefont {Altug}}, \ and\ \bibinfo {author} {\bibfnamefont {Y.}~\bibnamefont {Kivshar}},\ }\href@noop {} {\bibfield  {journal} {\bibinfo  {journal} {Advanced Materials}\ }\textbf {\bibinfo {volume} {36}},\ \bibinfo {pages} {2307494} (\bibinfo {year} {2024})}\BibitemShut {NoStop}%
\bibitem [{\citenamefont {Liu}\ \emph {et~al.}(2018{\natexlab{b}})\citenamefont {Liu}, \citenamefont {Vabishchevich}, \citenamefont {Vaskin}, \citenamefont {Reno}, \citenamefont {Keeler}, \citenamefont {Sinclair}, \citenamefont {Staude},\ and\ \citenamefont {Brener}}]{liu2018all}%
  \BibitemOpen
  \bibfield  {author} {\bibinfo {author} {\bibfnamefont {S.}~\bibnamefont {Liu}}, \bibinfo {author} {\bibfnamefont {P.~P.}\ \bibnamefont {Vabishchevich}}, \bibinfo {author} {\bibfnamefont {A.}~\bibnamefont {Vaskin}}, \bibinfo {author} {\bibfnamefont {J.~L.}\ \bibnamefont {Reno}}, \bibinfo {author} {\bibfnamefont {G.~A.}\ \bibnamefont {Keeler}}, \bibinfo {author} {\bibfnamefont {M.~B.}\ \bibnamefont {Sinclair}}, \bibinfo {author} {\bibfnamefont {I.}~\bibnamefont {Staude}}, \ and\ \bibinfo {author} {\bibfnamefont {I.}~\bibnamefont {Brener}},\ }\href@noop {} {\bibfield  {journal} {\bibinfo  {journal} {Nature communications}\ }\textbf {\bibinfo {volume} {9}},\ \bibinfo {pages} {2507} (\bibinfo {year} {2018}{\natexlab{b}})}\BibitemShut {NoStop}%
\bibitem [{\citenamefont {Timbrell}\ \emph {et~al.}(2018)\citenamefont {Timbrell}, \citenamefont {You}, \citenamefont {Kivshar},\ and\ \citenamefont {Panoiu}}]{timbrell2018comparative}%
  \BibitemOpen
  \bibfield  {author} {\bibinfo {author} {\bibfnamefont {D.}~\bibnamefont {Timbrell}}, \bibinfo {author} {\bibfnamefont {J.~W.}\ \bibnamefont {You}}, \bibinfo {author} {\bibfnamefont {Y.~S.}\ \bibnamefont {Kivshar}}, \ and\ \bibinfo {author} {\bibfnamefont {N.~C.}\ \bibnamefont {Panoiu}},\ }\href@noop {} {\bibfield  {journal} {\bibinfo  {journal} {Scientific Reports}\ }\textbf {\bibinfo {volume} {8}},\ \bibinfo {pages} {3586} (\bibinfo {year} {2018})}\BibitemShut {NoStop}%
\bibitem [{\citenamefont {Bouhelier}\ \emph {et~al.}(2003)\citenamefont {Bouhelier}, \citenamefont {Beversluis}, \citenamefont {Hartschuh},\ and\ \citenamefont {Novotny}}]{bouhelier2003near}%
  \BibitemOpen
  \bibfield  {author} {\bibinfo {author} {\bibfnamefont {A.}~\bibnamefont {Bouhelier}}, \bibinfo {author} {\bibfnamefont {M.}~\bibnamefont {Beversluis}}, \bibinfo {author} {\bibfnamefont {A.}~\bibnamefont {Hartschuh}}, \ and\ \bibinfo {author} {\bibfnamefont {L.}~\bibnamefont {Novotny}},\ }\href@noop {} {\bibfield  {journal} {\bibinfo  {journal} {Physical review letters}\ }\textbf {\bibinfo {volume} {90}},\ \bibinfo {pages} {013903} (\bibinfo {year} {2003})}\BibitemShut {NoStop}%
\bibitem [{\citenamefont {Timurdogan}\ \emph {et~al.}(2017)\citenamefont {Timurdogan}, \citenamefont {Poulton}, \citenamefont {Byrd},\ and\ \citenamefont {Watts}}]{timurdogan2017electric}%
  \BibitemOpen
  \bibfield  {author} {\bibinfo {author} {\bibfnamefont {E.}~\bibnamefont {Timurdogan}}, \bibinfo {author} {\bibfnamefont {C.~V.}\ \bibnamefont {Poulton}}, \bibinfo {author} {\bibfnamefont {M.}~\bibnamefont {Byrd}}, \ and\ \bibinfo {author} {\bibfnamefont {M.}~\bibnamefont {Watts}},\ }\href@noop {} {\bibfield  {journal} {\bibinfo  {journal} {Nature Photonics}\ }\textbf {\bibinfo {volume} {11}},\ \bibinfo {pages} {200} (\bibinfo {year} {2017})}\BibitemShut {NoStop}%
\bibitem [{\citenamefont {Lee}\ \emph {et~al.}(2019)\citenamefont {Lee}, \citenamefont {Taghinejad}, \citenamefont {Yan}, \citenamefont {Kim}, \citenamefont {Raju}, \citenamefont {Brown},\ and\ \citenamefont {Cai}}]{lee2019electrically}%
  \BibitemOpen
  \bibfield  {author} {\bibinfo {author} {\bibfnamefont {K.-T.}\ \bibnamefont {Lee}}, \bibinfo {author} {\bibfnamefont {M.}~\bibnamefont {Taghinejad}}, \bibinfo {author} {\bibfnamefont {J.}~\bibnamefont {Yan}}, \bibinfo {author} {\bibfnamefont {A.~S.}\ \bibnamefont {Kim}}, \bibinfo {author} {\bibfnamefont {L.}~\bibnamefont {Raju}}, \bibinfo {author} {\bibfnamefont {D.~K.}\ \bibnamefont {Brown}}, \ and\ \bibinfo {author} {\bibfnamefont {W.}~\bibnamefont {Cai}},\ }\href@noop {} {\bibfield  {journal} {\bibinfo  {journal} {ACS Photonics}\ }\textbf {\bibinfo {volume} {6}},\ \bibinfo {pages} {2663} (\bibinfo {year} {2019})}\BibitemShut {NoStop}%
\bibitem [{\citenamefont {Schriever}\ \emph {et~al.}(2010)\citenamefont {Schriever}, \citenamefont {Bohley},\ and\ \citenamefont {Wehrspohn}}]{schriever2010strain}%
  \BibitemOpen
  \bibfield  {author} {\bibinfo {author} {\bibfnamefont {C.}~\bibnamefont {Schriever}}, \bibinfo {author} {\bibfnamefont {C.}~\bibnamefont {Bohley}}, \ and\ \bibinfo {author} {\bibfnamefont {R.~B.}\ \bibnamefont {Wehrspohn}},\ }\href@noop {} {\bibfield  {journal} {\bibinfo  {journal} {Optics Letters}\ }\textbf {\bibinfo {volume} {35}},\ \bibinfo {pages} {273} (\bibinfo {year} {2010})}\BibitemShut {NoStop}%
\bibitem [{\citenamefont {Cazzanelli}\ \emph {et~al.}(2012)\citenamefont {Cazzanelli}, \citenamefont {Bianco}, \citenamefont {Borga}, \citenamefont {Pucker}, \citenamefont {Ghulinyan}, \citenamefont {Degoli}, \citenamefont {Luppi}, \citenamefont {V{\'e}niard}, \citenamefont {Ossicini}, \citenamefont {Modotto} \emph {et~al.}}]{cazzanelli2012second}%
  \BibitemOpen
  \bibfield  {author} {\bibinfo {author} {\bibfnamefont {M.}~\bibnamefont {Cazzanelli}}, \bibinfo {author} {\bibfnamefont {F.}~\bibnamefont {Bianco}}, \bibinfo {author} {\bibfnamefont {E.}~\bibnamefont {Borga}}, \bibinfo {author} {\bibfnamefont {G.}~\bibnamefont {Pucker}}, \bibinfo {author} {\bibfnamefont {M.}~\bibnamefont {Ghulinyan}}, \bibinfo {author} {\bibfnamefont {E.}~\bibnamefont {Degoli}}, \bibinfo {author} {\bibfnamefont {E.}~\bibnamefont {Luppi}}, \bibinfo {author} {\bibfnamefont {V.}~\bibnamefont {V{\'e}niard}}, \bibinfo {author} {\bibfnamefont {S.}~\bibnamefont {Ossicini}}, \bibinfo {author} {\bibfnamefont {D.}~\bibnamefont {Modotto}},  \emph {et~al.},\ }\href@noop {} {\bibfield  {journal} {\bibinfo  {journal} {Nature Materials}\ }\textbf {\bibinfo {volume} {11}},\ \bibinfo {pages} {148} (\bibinfo {year} {2012})}\BibitemShut {NoStop}%
\bibitem [{\citenamefont {Zhao}\ \emph {et~al.}(2022)\citenamefont {Zhao}, \citenamefont {Jia}, \citenamefont {Wang}, \citenamefont {Dong}, \citenamefont {Fang}, \citenamefont {Yang},\ and\ \citenamefont {Sun}}]{zhao2022second}%
  \BibitemOpen
  \bibfield  {author} {\bibinfo {author} {\bibfnamefont {Y.}~\bibnamefont {Zhao}}, \bibinfo {author} {\bibfnamefont {W.}~\bibnamefont {Jia}}, \bibinfo {author} {\bibfnamefont {X.-J.}\ \bibnamefont {Wang}}, \bibinfo {author} {\bibfnamefont {Y.}~\bibnamefont {Dong}}, \bibinfo {author} {\bibfnamefont {H.-H.}\ \bibnamefont {Fang}}, \bibinfo {author} {\bibfnamefont {Y.}~\bibnamefont {Yang}}, \ and\ \bibinfo {author} {\bibfnamefont {H.-B.}\ \bibnamefont {Sun}},\ }\href@noop {} {\bibfield  {journal} {\bibinfo  {journal} {Advanced Photonics Research}\ }\textbf {\bibinfo {volume} {3}},\ \bibinfo {pages} {2200157} (\bibinfo {year} {2022})}\BibitemShut {NoStop}%
\bibitem [{\citenamefont {Guo}\ \emph {et~al.}(2019)\citenamefont {Guo}, \citenamefont {Li},\ and\ \citenamefont {Guo}}]{guo2019enhanced}%
  \BibitemOpen
  \bibfield  {author} {\bibinfo {author} {\bibfnamefont {Z.}~\bibnamefont {Guo}}, \bibinfo {author} {\bibfnamefont {Z.}~\bibnamefont {Li}}, \ and\ \bibinfo {author} {\bibfnamefont {K.}~\bibnamefont {Guo}},\ }\href@noop {} {\bibfield  {journal} {\bibinfo  {journal} {Annalen der Physik}\ }\textbf {\bibinfo {volume} {531}},\ \bibinfo {pages} {1800470} (\bibinfo {year} {2019})}\BibitemShut {NoStop}%
\bibitem [{\citenamefont {Yang}\ \emph {et~al.}(2021)\citenamefont {Yang}, \citenamefont {Li}, \citenamefont {Kang}, \citenamefont {Guo}, \citenamefont {Zhang},\ and\ \citenamefont {Guo}}]{yang2021enhanced}%
  \BibitemOpen
  \bibfield  {author} {\bibinfo {author} {\bibfnamefont {G.}~\bibnamefont {Yang}}, \bibinfo {author} {\bibfnamefont {Z.}~\bibnamefont {Li}}, \bibinfo {author} {\bibfnamefont {Q.}~\bibnamefont {Kang}}, \bibinfo {author} {\bibfnamefont {K.}~\bibnamefont {Guo}}, \bibinfo {author} {\bibfnamefont {H.}~\bibnamefont {Zhang}}, \ and\ \bibinfo {author} {\bibfnamefont {Z.}~\bibnamefont {Guo}},\ }\href@noop {} {\bibfield  {journal} {\bibinfo  {journal} {Journal of Physics D: Applied Physics}\ }\textbf {\bibinfo {volume} {54}},\ \bibinfo {pages} {175110} (\bibinfo {year} {2021})}\BibitemShut {NoStop}%
\bibitem [{\citenamefont {Scalora}\ \emph {et~al.}(2019)\citenamefont {Scalora}, \citenamefont {Trull}, \citenamefont {Cojocaru}, \citenamefont {Vincenti}, \citenamefont {Carletti}, \citenamefont {de~Ceglia}, \citenamefont {Akozbek},\ and\ \citenamefont {De~Angelis}}]{scalora2019resonant}%
  \BibitemOpen
  \bibfield  {author} {\bibinfo {author} {\bibfnamefont {M.}~\bibnamefont {Scalora}}, \bibinfo {author} {\bibfnamefont {J.}~\bibnamefont {Trull}}, \bibinfo {author} {\bibfnamefont {C.}~\bibnamefont {Cojocaru}}, \bibinfo {author} {\bibfnamefont {M.}~\bibnamefont {Vincenti}}, \bibinfo {author} {\bibfnamefont {L.}~\bibnamefont {Carletti}}, \bibinfo {author} {\bibfnamefont {D.}~\bibnamefont {de~Ceglia}}, \bibinfo {author} {\bibfnamefont {N.}~\bibnamefont {Akozbek}}, \ and\ \bibinfo {author} {\bibfnamefont {C.}~\bibnamefont {De~Angelis}},\ }\href@noop {} {\bibfield  {journal} {\bibinfo  {journal} {J. Opt. Soc. Am. B}\ }\textbf {\bibinfo {volume} {36}},\ \bibinfo {pages} {2346} (\bibinfo {year} {2019})}\BibitemShut {NoStop}%
\bibitem [{\citenamefont {Hallman}\ \emph {et~al.}(2023)\citenamefont {Hallman}, \citenamefont {Rodr{\'\i}guez-Sun{\'e}}, \citenamefont {Trull}, \citenamefont {Cojocaru}, \citenamefont {Vincenti}, \citenamefont {Akozbek}, \citenamefont {Vilaseca},\ and\ \citenamefont {Scalora}}]{hallman2023harmonic}%
  \BibitemOpen
  \bibfield  {author} {\bibinfo {author} {\bibfnamefont {K.}~\bibnamefont {Hallman}}, \bibinfo {author} {\bibfnamefont {L.}~\bibnamefont {Rodr{\'\i}guez-Sun{\'e}}}, \bibinfo {author} {\bibfnamefont {J.}~\bibnamefont {Trull}}, \bibinfo {author} {\bibfnamefont {C.}~\bibnamefont {Cojocaru}}, \bibinfo {author} {\bibfnamefont {M.~A.}\ \bibnamefont {Vincenti}}, \bibinfo {author} {\bibfnamefont {N.}~\bibnamefont {Akozbek}}, \bibinfo {author} {\bibfnamefont {R.}~\bibnamefont {Vilaseca}}, \ and\ \bibinfo {author} {\bibfnamefont {M.}~\bibnamefont {Scalora}},\ }\href@noop {} {\bibfield  {journal} {\bibinfo  {journal} {Optics Express}\ }\textbf {\bibinfo {volume} {31}},\ \bibinfo {pages} {792} (\bibinfo {year} {2023})}\BibitemShut {NoStop}%
\bibitem [{\citenamefont {Makarov}\ \emph {et~al.}(2017)\citenamefont {Makarov}, \citenamefont {Petrov}, \citenamefont {Zywietz}, \citenamefont {Milichko}, \citenamefont {Zuev}, \citenamefont {Lopanitsyna}, \citenamefont {Kuksin}, \citenamefont {Mukhin}, \citenamefont {Zograf}, \citenamefont {Ubyivovk} \emph {et~al.}}]{makarov2017efficient}%
  \BibitemOpen
  \bibfield  {author} {\bibinfo {author} {\bibfnamefont {S.~V.}\ \bibnamefont {Makarov}}, \bibinfo {author} {\bibfnamefont {M.~I.}\ \bibnamefont {Petrov}}, \bibinfo {author} {\bibfnamefont {U.}~\bibnamefont {Zywietz}}, \bibinfo {author} {\bibfnamefont {V.}~\bibnamefont {Milichko}}, \bibinfo {author} {\bibfnamefont {D.}~\bibnamefont {Zuev}}, \bibinfo {author} {\bibfnamefont {N.}~\bibnamefont {Lopanitsyna}}, \bibinfo {author} {\bibfnamefont {A.}~\bibnamefont {Kuksin}}, \bibinfo {author} {\bibfnamefont {I.}~\bibnamefont {Mukhin}}, \bibinfo {author} {\bibfnamefont {G.}~\bibnamefont {Zograf}}, \bibinfo {author} {\bibfnamefont {E.}~\bibnamefont {Ubyivovk}},  \emph {et~al.},\ }\href@noop {} {\bibfield  {journal} {\bibinfo  {journal} {Nano letters}\ }\textbf {\bibinfo {volume} {17}},\ \bibinfo {pages} {3047} (\bibinfo {year} {2017})}\BibitemShut {NoStop}%
\bibitem [{\citenamefont {Wiecha}\ \emph {et~al.}(2015)\citenamefont {Wiecha}, \citenamefont {Arbouet}, \citenamefont {Kallel}, \citenamefont {Periwal}, \citenamefont {Baron},\ and\ \citenamefont {Paillard}}]{wiecha2015enhanced}%
  \BibitemOpen
  \bibfield  {author} {\bibinfo {author} {\bibfnamefont {P.~R.}\ \bibnamefont {Wiecha}}, \bibinfo {author} {\bibfnamefont {A.}~\bibnamefont {Arbouet}}, \bibinfo {author} {\bibfnamefont {H.}~\bibnamefont {Kallel}}, \bibinfo {author} {\bibfnamefont {P.}~\bibnamefont {Periwal}}, \bibinfo {author} {\bibfnamefont {T.}~\bibnamefont {Baron}}, \ and\ \bibinfo {author} {\bibfnamefont {V.}~\bibnamefont {Paillard}},\ }\href@noop {} {\bibfield  {journal} {\bibinfo  {journal} {Physical Review B}\ }\textbf {\bibinfo {volume} {91}},\ \bibinfo {pages} {121416} (\bibinfo {year} {2015})}\BibitemShut {NoStop}%
\bibitem [{\citenamefont {Wiecha}\ \emph {et~al.}(2016)\citenamefont {Wiecha}, \citenamefont {Arbouet}, \citenamefont {Girard}, \citenamefont {Baron},\ and\ \citenamefont {Paillard}}]{wiecha2016origin}%
  \BibitemOpen
  \bibfield  {author} {\bibinfo {author} {\bibfnamefont {P.~R.}\ \bibnamefont {Wiecha}}, \bibinfo {author} {\bibfnamefont {A.}~\bibnamefont {Arbouet}}, \bibinfo {author} {\bibfnamefont {C.}~\bibnamefont {Girard}}, \bibinfo {author} {\bibfnamefont {T.}~\bibnamefont {Baron}}, \ and\ \bibinfo {author} {\bibfnamefont {V.}~\bibnamefont {Paillard}},\ }\href@noop {} {\bibfield  {journal} {\bibinfo  {journal} {Physical Review B}\ }\textbf {\bibinfo {volume} {93}},\ \bibinfo {pages} {125421} (\bibinfo {year} {2016})}\BibitemShut {NoStop}%
\bibitem [{\citenamefont {Galli}\ \emph {et~al.}(2010)\citenamefont {Galli}, \citenamefont {Gerace}, \citenamefont {Welna}, \citenamefont {Krauss}, \citenamefont {O’Faolain}, \citenamefont {Guizzetti},\ and\ \citenamefont {Andreani}}]{galli2010low}%
  \BibitemOpen
  \bibfield  {author} {\bibinfo {author} {\bibfnamefont {M.}~\bibnamefont {Galli}}, \bibinfo {author} {\bibfnamefont {D.}~\bibnamefont {Gerace}}, \bibinfo {author} {\bibfnamefont {K.}~\bibnamefont {Welna}}, \bibinfo {author} {\bibfnamefont {T.~F.}\ \bibnamefont {Krauss}}, \bibinfo {author} {\bibfnamefont {L.}~\bibnamefont {O’Faolain}}, \bibinfo {author} {\bibfnamefont {G.}~\bibnamefont {Guizzetti}}, \ and\ \bibinfo {author} {\bibfnamefont {L.~C.}\ \bibnamefont {Andreani}},\ }\href@noop {} {\bibfield  {journal} {\bibinfo  {journal} {Optics Express}\ }\textbf {\bibinfo {volume} {18}},\ \bibinfo {pages} {26613} (\bibinfo {year} {2010})}\BibitemShut {NoStop}%
\bibitem [{\citenamefont {Bar-David}\ and\ \citenamefont {Levy}(2019)}]{bar2019nonlinear}%
  \BibitemOpen
  \bibfield  {author} {\bibinfo {author} {\bibfnamefont {J.}~\bibnamefont {Bar-David}}\ and\ \bibinfo {author} {\bibfnamefont {U.}~\bibnamefont {Levy}},\ }\href@noop {} {\bibfield  {journal} {\bibinfo  {journal} {Nano Letters}\ }\textbf {\bibinfo {volume} {19}},\ \bibinfo {pages} {1044} (\bibinfo {year} {2019})}\BibitemShut {NoStop}%
\bibitem [{\citenamefont {Xu}\ \emph {et~al.}(2021)\citenamefont {Xu}, \citenamefont {Plum}, \citenamefont {Savinov},\ and\ \citenamefont {Zheludev}}]{xu2021second}%
  \BibitemOpen
  \bibfield  {author} {\bibinfo {author} {\bibfnamefont {J.}~\bibnamefont {Xu}}, \bibinfo {author} {\bibfnamefont {E.}~\bibnamefont {Plum}}, \bibinfo {author} {\bibfnamefont {V.}~\bibnamefont {Savinov}}, \ and\ \bibinfo {author} {\bibfnamefont {N.~I.}\ \bibnamefont {Zheludev}},\ }\href@noop {} {\bibfield  {journal} {\bibinfo  {journal} {APL Photonics}\ }\textbf {\bibinfo {volume} {6}} (\bibinfo {year} {2021})}\BibitemShut {NoStop}%
\bibitem [{\citenamefont {Wang}\ \emph {et~al.}(2024)\citenamefont {Wang}, \citenamefont {Tonkaev}, \citenamefont {Koshelev}, \citenamefont {Lai}, \citenamefont {Kruk}, \citenamefont {Song}, \citenamefont {Kivshar},\ and\ \citenamefont {Panoiu}}]{wang2024resonantly}%
  \BibitemOpen
  \bibfield  {author} {\bibinfo {author} {\bibfnamefont {J.~T.}\ \bibnamefont {Wang}}, \bibinfo {author} {\bibfnamefont {P.}~\bibnamefont {Tonkaev}}, \bibinfo {author} {\bibfnamefont {K.}~\bibnamefont {Koshelev}}, \bibinfo {author} {\bibfnamefont {F.}~\bibnamefont {Lai}}, \bibinfo {author} {\bibfnamefont {S.}~\bibnamefont {Kruk}}, \bibinfo {author} {\bibfnamefont {Q.}~\bibnamefont {Song}}, \bibinfo {author} {\bibfnamefont {Y.}~\bibnamefont {Kivshar}}, \ and\ \bibinfo {author} {\bibfnamefont {N.~C.}\ \bibnamefont {Panoiu}},\ }\href@noop {} {\bibfield  {journal} {\bibinfo  {journal} {Opto-Electronic Advances}\ ,\ \bibinfo {pages} {230186}} (\bibinfo {year} {2024})}\BibitemShut {NoStop}%
\bibitem [{\citenamefont {Guyot-Sionnest}\ \emph {et~al.}(1986)\citenamefont {Guyot-Sionnest}, \citenamefont {Chen},\ and\ \citenamefont {Shen}}]{shen1986}%
  \BibitemOpen
  \bibfield  {author} {\bibinfo {author} {\bibfnamefont {P.}~\bibnamefont {Guyot-Sionnest}}, \bibinfo {author} {\bibfnamefont {W.}~\bibnamefont {Chen}}, \ and\ \bibinfo {author} {\bibfnamefont {Y.}~\bibnamefont {Shen}},\ }\href@noop {} {\bibfield  {journal} {\bibinfo  {journal} {Phys. Rev. B}\ }\textbf {\bibinfo {volume} {33}},\ \bibinfo {pages} {8254} (\bibinfo {year} {1986})}\BibitemShut {NoStop}%
\bibitem [{\citenamefont {Sipe}\ \emph {et~al.}(1987)\citenamefont {Sipe}, \citenamefont {Moss}, ,\ and\ \citenamefont {van Driel}}]{sipe1987}%
  \BibitemOpen
  \bibfield  {author} {\bibinfo {author} {\bibfnamefont {J.}~\bibnamefont {Sipe}}, \bibinfo {author} {\bibfnamefont {D.}~\bibnamefont {Moss}}, , \ and\ \bibinfo {author} {\bibfnamefont {H.}~\bibnamefont {van Driel}},\ }\href@noop {} {\bibfield  {journal} {\bibinfo  {journal} {Phys. Rev. B}\ }\textbf {\bibinfo {volume} {35}},\ \bibinfo {pages} {1129} (\bibinfo {year} {1987})}\BibitemShut {NoStop}%
\bibitem [{\citenamefont {Shen}(1999)}]{shen1999}%
  \BibitemOpen
  \bibfield  {author} {\bibinfo {author} {\bibfnamefont {Y.}~\bibnamefont {Shen}},\ }\href@noop {} {\bibfield  {journal} {\bibinfo  {journal} {Appl. Phys. B}\ }\textbf {\bibinfo {volume} {68}},\ \bibinfo {pages} {295} (\bibinfo {year} {1999})}\BibitemShut {NoStop}%
\bibitem [{\citenamefont {Lee}\ and\ \citenamefont {Downer}(1998)}]{lee1998reflected}%
  \BibitemOpen
  \bibfield  {author} {\bibinfo {author} {\bibfnamefont {Y.-S.}\ \bibnamefont {Lee}}\ and\ \bibinfo {author} {\bibfnamefont {M.}~\bibnamefont {Downer}},\ }\href@noop {} {\bibfield  {journal} {\bibinfo  {journal} {Optics letters}\ }\textbf {\bibinfo {volume} {23}},\ \bibinfo {pages} {918} (\bibinfo {year} {1998})}\BibitemShut {NoStop}%
\bibitem [{\citenamefont {Li}(1980)}]{li1980refractive}%
  \BibitemOpen
  \bibfield  {author} {\bibinfo {author} {\bibfnamefont {H.}~\bibnamefont {Li}},\ }\href@noop {} {\bibfield  {journal} {\bibinfo  {journal} {Journal of Physical and Chemical Reference Data}\ }\textbf {\bibinfo {volume} {9}},\ \bibinfo {pages} {561} (\bibinfo {year} {1980})}\BibitemShut {NoStop}%
\bibitem [{not()}]{noteSI}%
  \BibitemOpen
  \href@noop {} {}\bibinfo {note} {See Supplemental Material for (Figure S1) transmission spectra for a-Si metasurfaces, (Figure S2) experimental setup schematics, (Figure S3) angles dependence of third harmonic generation on pump polarization, (Figure S4) power dependence of third harmonic from metasurface outside the resonance.}\BibitemShut {Stop}%
\bibitem [{\citenamefont {Fomenko}\ \emph {et~al.}(2001)\citenamefont {Fomenko}, \citenamefont {Lami},\ and\ \citenamefont {Borguet}}]{fomenko2001nonquadratic}%
  \BibitemOpen
  \bibfield  {author} {\bibinfo {author} {\bibfnamefont {V.}~\bibnamefont {Fomenko}}, \bibinfo {author} {\bibfnamefont {J.-F.}\ \bibnamefont {Lami}}, \ and\ \bibinfo {author} {\bibfnamefont {E.}~\bibnamefont {Borguet}},\ }\href@noop {} {\bibfield  {journal} {\bibinfo  {journal} {Physical Review B}\ }\textbf {\bibinfo {volume} {63}},\ \bibinfo {pages} {121316} (\bibinfo {year} {2001})}\BibitemShut {NoStop}%
\bibitem [{\citenamefont {Wang}\ \emph {et~al.}(1998)\citenamefont {Wang}, \citenamefont {L{\"u}pke}, \citenamefont {Di~Ventra}, \citenamefont {Pantelides}, \citenamefont {Gilligan}, \citenamefont {Tolk}, \citenamefont {Kizilyalli}, \citenamefont {Roy}, \citenamefont {Margaritondo},\ and\ \citenamefont {Lucovsky}}]{wang1998coupled}%
  \BibitemOpen
  \bibfield  {author} {\bibinfo {author} {\bibfnamefont {W.}~\bibnamefont {Wang}}, \bibinfo {author} {\bibfnamefont {G.}~\bibnamefont {L{\"u}pke}}, \bibinfo {author} {\bibfnamefont {M.}~\bibnamefont {Di~Ventra}}, \bibinfo {author} {\bibfnamefont {S.}~\bibnamefont {Pantelides}}, \bibinfo {author} {\bibfnamefont {J.}~\bibnamefont {Gilligan}}, \bibinfo {author} {\bibfnamefont {N.}~\bibnamefont {Tolk}}, \bibinfo {author} {\bibfnamefont {I.}~\bibnamefont {Kizilyalli}}, \bibinfo {author} {\bibfnamefont {P.}~\bibnamefont {Roy}}, \bibinfo {author} {\bibfnamefont {G.}~\bibnamefont {Margaritondo}}, \ and\ \bibinfo {author} {\bibfnamefont {G.}~\bibnamefont {Lucovsky}},\ }\href@noop {} {\bibfield  {journal} {\bibinfo  {journal} {Physical Review Letters}\ }\textbf {\bibinfo {volume} {81}},\ \bibinfo {pages} {4224} (\bibinfo {year} {1998})}\BibitemShut {NoStop}%
\bibitem [{\citenamefont {Wen}\ \emph {et~al.}(2018)\citenamefont {Wen}, \citenamefont {Xu}, \citenamefont {Zhao}, \citenamefont {Khurgin},\ and\ \citenamefont {Xiong}}]{wen2018plasmonic}%
  \BibitemOpen
  \bibfield  {author} {\bibinfo {author} {\bibfnamefont {X.}~\bibnamefont {Wen}}, \bibinfo {author} {\bibfnamefont {W.}~\bibnamefont {Xu}}, \bibinfo {author} {\bibfnamefont {W.}~\bibnamefont {Zhao}}, \bibinfo {author} {\bibfnamefont {J.~B.}\ \bibnamefont {Khurgin}}, \ and\ \bibinfo {author} {\bibfnamefont {Q.}~\bibnamefont {Xiong}},\ }\href@noop {} {\bibfield  {journal} {\bibinfo  {journal} {Nano Letters}\ }\textbf {\bibinfo {volume} {18}},\ \bibinfo {pages} {1686} (\bibinfo {year} {2018})}\BibitemShut {NoStop}%
\bibitem [{\citenamefont {Sun}\ \emph {et~al.}(2023)\citenamefont {Sun}, \citenamefont {Larin}, \citenamefont {Mozharov}, \citenamefont {Ageev}, \citenamefont {Pashina}, \citenamefont {Komissarenko}, \citenamefont {Mukhin}, \citenamefont {Petrov}, \citenamefont {Makarov}, \citenamefont {Belov} \emph {et~al.}}]{sun2023all}%
  \BibitemOpen
  \bibfield  {author} {\bibinfo {author} {\bibfnamefont {Y.}~\bibnamefont {Sun}}, \bibinfo {author} {\bibfnamefont {A.}~\bibnamefont {Larin}}, \bibinfo {author} {\bibfnamefont {A.}~\bibnamefont {Mozharov}}, \bibinfo {author} {\bibfnamefont {E.}~\bibnamefont {Ageev}}, \bibinfo {author} {\bibfnamefont {O.}~\bibnamefont {Pashina}}, \bibinfo {author} {\bibfnamefont {F.}~\bibnamefont {Komissarenko}}, \bibinfo {author} {\bibfnamefont {I.}~\bibnamefont {Mukhin}}, \bibinfo {author} {\bibfnamefont {M.}~\bibnamefont {Petrov}}, \bibinfo {author} {\bibfnamefont {S.}~\bibnamefont {Makarov}}, \bibinfo {author} {\bibfnamefont {P.}~\bibnamefont {Belov}},  \emph {et~al.},\ }\href@noop {} {\bibfield  {journal} {\bibinfo  {journal} {Light: Science \& Applications}\ }\textbf {\bibinfo {volume} {12}},\ \bibinfo {pages} {237} (\bibinfo {year} {2023})}\BibitemShut {NoStop}%
\bibitem [{\citenamefont {Tancogne-Dejean}\ \emph {et~al.}(2016)\citenamefont {Tancogne-Dejean}, \citenamefont {Giorgetti},\ and\ \citenamefont {V{\'e}niard}}]{tancogne2016ab}%
  \BibitemOpen
  \bibfield  {author} {\bibinfo {author} {\bibfnamefont {N.}~\bibnamefont {Tancogne-Dejean}}, \bibinfo {author} {\bibfnamefont {C.}~\bibnamefont {Giorgetti}}, \ and\ \bibinfo {author} {\bibfnamefont {V.}~\bibnamefont {V{\'e}niard}},\ }\href@noop {} {\bibfield  {journal} {\bibinfo  {journal} {Physical Review B}\ }\textbf {\bibinfo {volume} {94}},\ \bibinfo {pages} {125301} (\bibinfo {year} {2016})}\BibitemShut {NoStop}%
\bibitem [{\citenamefont {Apostolova}\ and\ \citenamefont {Obreshkov}(2021)}]{apostolova2021high}%
  \BibitemOpen
  \bibfield  {author} {\bibinfo {author} {\bibfnamefont {T.}~\bibnamefont {Apostolova}}\ and\ \bibinfo {author} {\bibfnamefont {B.}~\bibnamefont {Obreshkov}},\ }\href@noop {} {\bibfield  {journal} {\bibinfo  {journal} {The European Physical Journal D}\ }\textbf {\bibinfo {volume} {75}},\ \bibinfo {pages} {1} (\bibinfo {year} {2021})}\BibitemShut {NoStop}%
\bibitem [{\citenamefont {Ovchinnikov}\ \emph {et~al.}(2019)\citenamefont {Ovchinnikov}, \citenamefont {Chefonov}, \citenamefont {Mishina},\ and\ \citenamefont {Agranat}}]{ovchinnikov2019second}%
  \BibitemOpen
  \bibfield  {author} {\bibinfo {author} {\bibfnamefont {A.}~\bibnamefont {Ovchinnikov}}, \bibinfo {author} {\bibfnamefont {O.}~\bibnamefont {Chefonov}}, \bibinfo {author} {\bibfnamefont {E.}~\bibnamefont {Mishina}}, \ and\ \bibinfo {author} {\bibfnamefont {M.}~\bibnamefont {Agranat}},\ }\href@noop {} {\bibfield  {journal} {\bibinfo  {journal} {Scientific Reports}\ }\textbf {\bibinfo {volume} {9}},\ \bibinfo {pages} {9753} (\bibinfo {year} {2019})}\BibitemShut {NoStop}%
\bibitem [{\citenamefont {Ciappina}\ \emph {et~al.}(2012)\citenamefont {Ciappina}, \citenamefont {Biegert}, \citenamefont {Quidant},\ and\ \citenamefont {Lewenstein}}]{ciappina2012high}%
  \BibitemOpen
  \bibfield  {author} {\bibinfo {author} {\bibfnamefont {M.~F.}\ \bibnamefont {Ciappina}}, \bibinfo {author} {\bibfnamefont {J.}~\bibnamefont {Biegert}}, \bibinfo {author} {\bibfnamefont {R.}~\bibnamefont {Quidant}}, \ and\ \bibinfo {author} {\bibfnamefont {M.}~\bibnamefont {Lewenstein}},\ }\href@noop {} {\bibfield  {journal} {\bibinfo  {journal} {Physical Review A}\ }\textbf {\bibinfo {volume} {85}},\ \bibinfo {pages} {033828} (\bibinfo {year} {2012})}\BibitemShut {NoStop}%
\bibitem [{\citenamefont {Mukhopadhyay}\ \emph {et~al.}(2023)\citenamefont {Mukhopadhyay}, \citenamefont {Rodriguez-Sun{\'e}}, \citenamefont {Cojocaru}, \citenamefont {Vincenti}, \citenamefont {Hallman}, \citenamefont {Leo}, \citenamefont {Belchovski}, \citenamefont {de~Ceglia}, \citenamefont {Scalora},\ and\ \citenamefont {Trull}}]{mukhopadhyay2023three}%
  \BibitemOpen
  \bibfield  {author} {\bibinfo {author} {\bibfnamefont {S.}~\bibnamefont {Mukhopadhyay}}, \bibinfo {author} {\bibfnamefont {L.}~\bibnamefont {Rodriguez-Sun{\'e}}}, \bibinfo {author} {\bibfnamefont {C.}~\bibnamefont {Cojocaru}}, \bibinfo {author} {\bibfnamefont {M.}~\bibnamefont {Vincenti}}, \bibinfo {author} {\bibfnamefont {K.}~\bibnamefont {Hallman}}, \bibinfo {author} {\bibfnamefont {G.}~\bibnamefont {Leo}}, \bibinfo {author} {\bibfnamefont {M.}~\bibnamefont {Belchovski}}, \bibinfo {author} {\bibfnamefont {D.}~\bibnamefont {de~Ceglia}}, \bibinfo {author} {\bibfnamefont {M.}~\bibnamefont {Scalora}}, \ and\ \bibinfo {author} {\bibfnamefont {J.}~\bibnamefont {Trull}},\ }\href@noop {} {\bibfield  {journal} {\bibinfo  {journal} {APL Photonics}\ }\textbf {\bibinfo {volume} {8}} (\bibinfo {year} {2023})}\BibitemShut {NoStop}%
\end{thebibliography}
\end{document}